\documentclass[12pt]{iopart}

\begin{document}
\title{General flux to a trap in one and three dimensions}

\author {Robert M. Ziff,$^{1}$ Satya N. Majumdar,$^{2}$ Alain Comtet$^{2,3}$ }
\address{
{\small $^1$ Michigan Center for
Theoretical Physics and Department of Chemical
Engineering, University of Michigan, Ann Arbor, MI 48109-2136, USA}\\
{\small $^2$ Laboratoire de Physique Th\'eorique et Mod\`eles Statistiques,
        Universit\'e Paris-Sud, B\^at.\ 100, 91405 Orsay Cedex, France}\\
{\small $^3$ Institut Henri Poincar\'e, 11 rue Pierre et Marie Curie, 75005 Paris, France}
}
\ead{rziff@umich.edu, majumdar@lptms.u-psud.fr, comtet@ipno.in2p3.fr}

\date{\today}

\begin{abstract}
The problem of the flux to a spherical trap in one and three dimensions, for diffusing
particles undergoing discrete-time jumps with a given radial probability distribution,
is solved in general, verifying the Smoluchowski-like solution in which the
effective trap radius is reduced by an amount proportional to the
jump length.  This reduction in the effective trap radius corresponds to
the Milne extrapolation length.
 \bigskip
\end{abstract}
\maketitle

\section{Introduction}

A classical problem in diffusion and reaction theory is the flux of particles, initially
uniform in space, to a trap.  When the particles are moving by simple diffusion,
this flux can be found by solving the diffusion equation with an adsorbing boundary
at the surface of the trap. For example, for a system with a spherical trap
of radius $R$, the time-dependent density $\rho(\vec r,t)$ evolves via the diffusion
equation, $\partial_t \rho =D\nabla^2 \rho$ ($D$ being the diffusion coefficient), starting from the 
initial condition
$\rho(\vec r, 0)=\rho_0$ for $r>R$ and $\rho(\vec r, 0)=0$ for $r\le R$. The diffusion
equation is subject to the boundary conditions, $\rho(\vec r,t)=0$ for $r=R$ 
and $\rho(\vec r, t)\to \rho_0$ as 
$r\to \infty$, for all time $t$. For spatial dimensions $d>2$, the density profile 
outside
the sphere approaches a steady state as $t\to \infty$, reflecting the fact that Brownian motion
is transient for $d>2$~\cite{Redner}. For example, in three dimensions, the full time-dependent density 
profile 
(which is spherically symmetric) can be explicitly obtained for all $t$,
\begin{equation}
\rho(r,t) = \frac{\rho_0}{r}\left[r-R\, {\rm erfc}\left(\frac{r-R}{\sqrt{4Dt}}\right)\right]\quad\quad 
{\rm for}\quad r\ge R \ ,
\label{denprof1}
\end{equation}
where ${\rm erfc}(z) = (2/\sqrt{\pi}) \int_z^{\infty} e^{-u^2} du$.
In particular, as $t\to \infty$, the density approaches a stationary profile outside the sphere,
\begin{equation}
\rho(r, \infty)= \frac{\rho_0}{r}(r-R) \ .
\label{denprof2}
\end{equation}
The flux $\Phi(t)=4\pi R^2 D ({\partial \rho}/{\partial r})|_{r=R}$, defined as the number of 
particles falling into the trap per unit time,
can be easily obtained from $\rho(r,t)$ in Eq.\ (\ref{denprof1}) and one gets the classical
result~\cite{Smolu,Chandra},
\begin{equation}
\Phi(t) = 4 \pi R D \rho_0 \left[ 1 + \frac{R}{\sqrt{\pi D t}} \right].
\label{classicalflux}
\end{equation}

On the other hand, when the particles are moving by discrete jumps at every time step $\tau$
(as opposed to undergoing  continuous-time Brownian motion),
the above results for the stationary density profile
in Eq.\ (\ref{denprof2}) as well as that for the flux in Eq.\ (\ref{classicalflux}) become 
modified.  In Ref.\ \cite{Ziff1} this discrete-jump problem was studied by 
an iterative numerical scheme, for particles undergoing fixed-length jumps of length $\ell$
and random angles (the so-called Rayleigh flights).
It was found that in $d=3$, while the density profile $\rho_n(r)$ after $n$ time steps
approaches a stationary limit as $n\to \infty$, the stationary profile $\rho_{\infty}(r)$ is  
different from
its continuous-time counterpart in Eq.\ (\ref{denprof2}). Very far from the surface of the sphere,
the stationary profile behaves as
\begin{equation}
\rho_{\infty}(r) \approx \frac{\rho_0}{r}\left[r- (R-\epsilon)\right]\quad\quad {\rm for}\quad r \gg R \ ,
\label{denprofinf}
\end{equation}
where $\epsilon = c\,{\ell}$ and the dimensionless constant $c=0.29795219\dots$ was determined numerically~\cite{Ziff1}. 
On the other hand, on the surface of the sphere, the stationary density profile approaches
a positive value
\begin{equation}
\rho_{\infty}(R) \approx 0.408245\, \frac{\rho_0 \ell}{R} \ ,
\label{denprofR}
\end{equation}
where the constant $0.408245$ was evaluated numerically in Ref.\ \cite{Ziff1}.
This is in marked contrast to the continuous-time Brownian case where the stationary density profile 
vanishes on the surface of the sphere. The distance $\epsilon = c\,\ell$ is 
the `Milne extrapolation length'~\cite{BT,CZ,Williams} that represents the distance inside the surface 
where the far steady-state 
solution in Eq.\ (\ref{denprofinf}) extrapolates to zero.  
The same length $\epsilon$ also appears in the expression for flux~\cite{Ziff1}
\begin{equation}
\Phi(t) \approx 4 \pi D \rho_0 (R - \epsilon) \left[ 1 + \frac{ R - \epsilon}{\sqrt{\pi D t}} + 
{\cal O} 
(t^{-3/2})\right] \ ,
\label{rzflux}
\end{equation}
for large $t$. This result was obtained numerically in Ref.\ \cite{Ziff1}.

Recently we have rigorously proven the result in Eq.\ (\ref{denprofinf}) for the large-distance
stationary profile and found an analytical expression for $c$~\cite{MCZ}:
\begin{equation}
c = - \frac{1}{\pi}\, \int_0^{\infty} \frac{dk}{k^2}\, \ln \left[\frac{6}{k^2}\left(1-\frac{\sin
k}{k}\right)\right] = 0.29795219028\dots \ .
\label{c2}
\end{equation}
Remarkably, the same constant also appeared in the leading finite-size correction 
to the expected maximum of a discrete-time random walker moving on a continuous 
line where at each time step the walker jumps by a distance $\xi$ that is chosen from a
uniform distribution~\cite{CFFH,CM}. In Ref.\ \cite{CFFH}, the constant was evaluated 
numerically by summing a rather complicated double infinite series. Later an exact expression
as in Eq.\ (\ref{c2}) was derived in Ref.\ \cite{CM}. In Ref.\ \cite{MCZ}, we showed 
why the same constant also appeared in the three-dimensional discrete flux problem.  
For the $3$-d flux problem, it was further proven~\cite{MCZ} that the
density on the surface of the 
sphere approaches a constant 
as in Eq.\ (\ref{denprofR}):
\begin{equation}
\rho_{\infty}(r)= \frac{1}{\sqrt{6}}\, \frac{\rho_0 \ell}{R}\ .
\label{denprofR2}
\end{equation}
Thus the constant $0.408245$ in Eq.\ (\ref{denprofR}), found numerically
in Ref.\ \cite{Ziff1}, was proven to be simply $1/\sqrt{6}$. Indeed, in Ref.\ \cite{MCZ},
an analytical expression for the full stationary density profile for all 
$r\ge R$ (or rather its Laplace transform) was obtained.

Having obtained the stationary density profile, the next objective is to prove
the expression for the time-dependent
flux in Eq.\ (\ref{rzflux}) that was found numerically in Ref.\ \cite{Ziff1}.
If one uses the form in Eq.\ (\ref{denprofinf}) of the steady-state solution far from the sphere
in the expression for the flux in the continuous-time diffusion problem, one gets~\cite{MCZ}
$\Phi = 4\pi r^2 D\, (d\rho_{\infty}(r)/dr) = 4\pi D \rho_0 (R-\epsilon)$,
 which is 
precisely the leading time-independent term in the expression for flux in Eq.\ (\ref{rzflux}).
Even though this reproduces correctly the leading term in the flux in Eq.\ (\ref{rzflux}), it 
is not entirely satisfactory since we are using a definition
of flux which is valid only for the continuous-time diffusion equation and not for the discrete-time 
jump process. Moreover, just the knowledge of the stationary density profile is not enough
to prove the leading time-dependent behaviour, namely the $1/\sqrt{t}$ decay to the stationary flux 
in Eq.\ (\ref{rzflux}). Thus, to rigorously prove the two leading terms in Eq.\ (\ref{rzflux}),
we need to obtain the time-dependent solution $\rho_n(r)$ of the discrete-time process for large 
$n$ and then use this solution in an appropriate expression for the discrete flux (see below).
This is precisely what is achieved in this paper.

The main achievements of this paper are: (i) to prove rigorously the result for the discrete
flux in Eq.\ (\ref{rzflux}) for the two leading terms and then (ii) to generalize 
the results in Eqs.\ (\ref{denprofinf}), (\ref{denprofR}) and (\ref{rzflux}) to the case where the
jump length $|\vec r-\vec r'|$ from $\vec r'$ to $\vec r$ is not of a fixed length
(as in Rayleigh flights) but is a random variable drawn from
a distribution $W(|\vec r-\vec r'|)$ that is bounded above (with an upper cut-off $< 2R$), 
but is otherwise arbitrary. The stationary density profile for the general case was already calculated
in detail in Ref.\ \cite{MCZ} and it was shown that Eq.\ (\ref{denprofinf}) 
is valid for the general case where  $\epsilon$ has an explicit expression~\cite{MCZ}
\begin{equation}
\epsilon = - \frac{1}{\pi}\, \int_0^{\infty} \frac{dk}{k^2}\, \ln \left[\frac{1-{\hat
f}(k)}{\sigma^2k^2/2}\right] \ ,
\label{stat5}
\end{equation}
where $\hat f(k) = \int_{-\infty}^\infty f(x) e^{i k x} dx$
is the Fourier transform of the function 
\begin{equation}
f(x) = 2\pi \int_{|x|}^{\infty} W(u)\,u\, du \ ,
 \label{jump}
 \end{equation}
and $\sigma^2= \int_{-\infty}^{\infty} f(x)\, x^2\, dx$ is the second moment of $f(x)$
which is assumed to be finite.  In terms of $W(u)$, we have
 $\sigma^2= (2 \pi / 3) \int_{-\infty}^{\infty} W(u)\, u^4\, du$.
For the case of the Rayleigh flight, the jump probability
$W(u) = \frac{1}{4\pi\, \ell^2}\delta(u - \ell)$ in three-dimensional space translates to
a uniform jump distribution for $f(x)$ \cite{CK},
\begin{equation}
f(x) =    \left\{ \begin{array}{l}
       \frac{1}{2 \ell}, \qquad |x| < \ell\\
        0,  \hbox{     otherwise} \ . \end{array} \right.
\label{uniform}
\end{equation}
Then $\hat f(k) = \sin k\ell / (k \ell)$, and
Eq.\ (\ref{stat5}) reduces to Eq.\ (\ref{c2}) with $\epsilon = c \ell$.

In addition, for the general jump distribution
$W(z)$, another remarkable universal result was proven~\cite{MCZ}. It was shown that
the stationary density on the surface of the sphere, 
properly rescaled by $R$ and $\sigma$,
is a universal dimensionless constant~\cite{MCZ}
\begin{equation}
\frac{\rho_{\infty}(R)}{\rho_0}\,\frac{R}{\sigma}=\frac{1}{\sqrt{2}} \ ,
\label{univ1}
\end{equation}
which generalizes Eq.\ (\ref{denprofR2}).
In this paper, we will prove that the asymptotic
expression for the discrete flux in Eq.\ (\ref{rzflux}) 
 is also valid for a general jump distribution $W(z)$
with the length $\epsilon$ given by the exact formula in Eq.\ (\ref{stat5}). 

Apart from the three-dimensional results mentioned above, in this paper we also study the 
one-dimensional discrete flux problem where the origin is a trap and initially
random walkers are distributed uniformly with density $\rho_0$ on the positive
semi-infinite line. At every discrete time step, each walker jumps independently by a distance $x$ 
that is chosen
from a normalized probability density function $f(x)$ with a finite second moment
$\sigma^2=\int_{-\infty}^{\infty} f(x)\, x^2\, dx$. Due to the recurrent nature of 
random walk
in one dimension, the density approaches zero everywhere on the positive line at late times
and the instantaneous flux to the origin also decays to zero with time. Our exact calculation 
shows that while the instantaneous flux decays algebraically as $1/\sqrt{n}$ for large time
$n$, the cumulative flux $\Phi_n^{\rm{cum}}$ (i.e., the total number of particles absorbed
by the origin up to $n$ steps) behaves asymptotically as
\begin{equation}
\Phi_n^{\rm{cum}} =\rho_0\,\left[\sigma \sqrt{\frac{2n}{\pi}}-\epsilon +  \frac{\mu_4 + 3 \sigma^4}{12 
\sigma^3} 
\sqrt{\frac1{2 \pi n}} +
{\cal O}(n^{-1})\right]  \ ,
\label{cumfd1}
\end{equation}
where $\mu_4= \int_{-\infty}^{\infty} f(x)\, x^4\, dx$ (assuming it is finite) and $\epsilon$ is the same 
quantity that appeared in the $3$-d flux problem in Eq.\ (\ref{stat5}).  The quantity $\epsilon$
however does not appear in the expression for the 1-d 
instantaneous flux as it does in three dimensions (at least up to ${\cal O}(n^{-3/2})$).   

The paper is organized as follows. In section II, we set up the basic integral equation 
satisfied by the time-dependent density profile for an arbitrary jump distribution in
three dimensions and show how one can reduce it to a one-dimensional integral equation.
We then briefly describe the exact solution to this integral equation first obtained
in Ref.\ \cite{MCZ}. In section III, we use this solution to calculate  $\Phi_n^{\rm{cum}}$
and from that derive the main result in Eq.\ (\ref{rzflux}).
In the process, we also prove analytically some other conjectures made in Ref.\ \cite{Ziff1}
on the basis of numerics. Section IV deals with the expressions for the flux in 
one dimension. Finally we conclude in section V with a summary and outlook.
Some of the details of the calculations are relegated to the Appendix.

\section{The Integral Equation for the density profile and its exact solution}

Let $\rho_n(\vec r)$ denote the density of particles at position $\vec r$ outside the
sphere of radius $R$ at time step $n$. Initially we have $\rho_0(\vec r) = \rho_0$
for all $r>R$. The density stays zero inside the sphere since it is a trap.
A particle at position $\vec r'$ (with $r'>R$) jumps in one time step
to a new position $\vec r$ and the jump distance $(\vec r-\vec r')$ is drawn
at each step independently from an isotropic distribution $W(|\vec r-\vec r'|)$ that
depends only on the magnitude of the jump length, but not on its direction.
The function $W(u)$ is normalized in the three-dimensional space, i.e.,
\begin{equation}
4\pi \int_0^{\infty} W(u) u^2 du=1\ .
\label{norm1}
\end{equation}
The time evolution of the density field $\rho_n(\vec r)$ is clearly Markovian
and is governed by the integral equation
\begin{equation}
\rho_n(\vec r) = \int_{r'\ge R} \rho_{n-1}(\vec r')\, W(|\vec r-\vec r'|)\, d\vec r' \ .
\label{evol1}
\end{equation}
The initial condition is spherically symmetric and the evolution equation preserves this
symmetry, hence $\rho_n(\vec r) = \rho_n(r)$ at all steps $n$. We then have
\begin{equation}
\rho_n(r)=  \int_{r'\ge R} \rho_{n-1}(r')\, W(|\vec r-\vec r'|)\, d\vec r' \ .
\label{evol2}
\end{equation}
Using $|\vec r-\vec r'|=\sqrt{r^2+r'^2 -2 r\,r'\, \cos{\theta}}$ 
where $\theta$ is the angle between $\vec r$ and $\vec r'$ and $d\vec r' = 2\pi r'^2 \sin {\theta} 
dr'\, d\theta$ in three dimensions we get
\begin{eqnarray}
\rho_n(r)&=& 2\pi \int_{R}^{\infty} \, dr' \rho_{n-1}(r') r'^2 \int_0^{\pi} 
W\left(\sqrt{r^2+r'^2-2rr'\cos{\theta}}\right) \sin{\theta} \, d\theta \nonumber \\
&=& \frac{2\pi}{r}\, \int_{R}^{\infty} dr' \,r'\, \rho_{n-1}(r')\, \int_{|r-r'|}^{r+r'} W(u)\, u\, du \ .
\label{evol3}
\end{eqnarray}
Let us define a new variable $F_n(r) = r \rho_n(r)/\rho_0$. Then $F_n(r)$ for $r\ge R$ evolves via
the one-dimensional integral equation
\begin{equation}
F_n(r) = 2\pi \int_{R}^{\infty} dr'\, F_{n-1}(r')\, \int_{|r-r'|}^{r+r'} W(u)\, u\, du
\quad {\hbox{with}}\quad
F_0(r)=r \ .
\label{evol4}
\end{equation}
A further simplification can be achieved by introducing a shift, i.e., defining
$z= r-R$ (distance from the surface of the sphere) and writing $F_n(r)= F_n(R+z)= H_n(z)$
where we have suppressed the $R$ dependence in $H_n(z)$ for convenience. Then $H_n(z)$
for $z\ge 0$ evolves via
\begin{equation}
H_n(z) = 2\pi \int_0^{\infty} dz'\, H_{n-1}(z')\, \int_{|z-z'|}^{z+z'+2R} W(u)\, u\, du \ ,
\label{evol5}
\end{equation}
with $H_0(z)= R+z$.

Eq.\ (\ref{evol5}) is very general and is valid for arbitrary isotropic jump distribution $W(|\vec r 
-\vec r'|)$. However, the reduced effective kernel in the one-dimensional integral equation 
(\ref{evol5}) --- namely $f(z,z') = 2\pi\int_{|z-z'|}^{z+z'+2R} W(u)\, u\, du$ --- is non-stationary
since it depends on both $r$ and $r'$ and not just on its difference $|r-r'|$. 
It is difficult to solve this integral equation with a non-stationary kernel. To simplify further,
we will henceforth assume that the jump distribution $W(z=|\vec r-\vec r'|)$ is bounded 
above, i.e., it has a finite support with an upper cut-off $z_{\rm max}$. This means
$W(z)=0$ for $z> z_{\rm max}$. If we further assume that $z_{\rm max}<2R$, it follows
that one can replace the upper limit of integration in $f(z,z')$ by $\infty$ and 
Eq.\ (\ref{evol5}) simplifies to one with a stationary kernel,
\begin{equation}
H_n(z)= \int_0^{\infty} dz' \, H_{n-1}(z')\, f(z-z')\, dz' \quad\quad {\rm with}\quad H_0(z)=R+z \ ,
\label{evol6}
\end{equation}
where $f(x)$ is a symmetric non-negative function
\begin{equation}
f(x)= 2\pi \int_{|x|}^{\infty} W(u)\, u \, du \ .
\label{kernel1}
\end{equation}
Moreover, it is easy to show from Eq.\ (\ref{kernel1}), via integration by parts and then using
Eq.\ (\ref{norm1}), that 
$f(x)$ is 
normalized to unity: $\int_{-\infty}^{\infty} f(x) dx=1$. Thus $f(x)$ can be regarded
as a probability density function. For later purposes, it would further be convenient to break
the solution of Eq.\ (\ref{evol6}) into two $R$-independent parts:
$H_n(z)= Q_n(z) + R q_n(z)$ where $Q_n(z)$ and $q_n(z)$ evolve by the same integral equation, albeit with 
different initial conditions:
\begin{eqnarray}
Q_n(z) &=& \int_0^{\infty} dz' \, Q_{n-1}(z')\, f(z-z')\, dz' \quad\quad {\rm with}\quad Q_0(z)=z\ , \label{Qn1l} 
\\
q_n(z) &=& \int_0^{\infty} dz' \, q_{n-1}(z')\, f(z-z')\, dz' \quad\quad {\rm with}\quad q_0(z)=1\ . \label{qn1}  
\end{eqnarray}
Once the solutions to these two equations are known, the density profile $\rho_n(r)$ can be 
obtained via
\begin{equation}
\rho_n(r) = \frac{\rho_0}{r}\, \left[Q_n(r-R)+R \,q_n(r-R)\right] \ .
\label{denprofn}
\end{equation}

The solutions to the two integral equations in Eqs.\ (\ref{Qn1l}) and (\ref{qn1}) were obtained
explicitly (in the Laplace domain) in Ref.\ \cite{MCZ}. We will not provide the derivation here, but just 
mention the main results. It was shown that the two following Laplace transforms
\begin{equation}   
\hat\psi_{\rm I}(u,s) = \int_0^\infty \left( \sum_{n=0}^\infty q_n(z) s^n\right) e^{-u z} dz
\label{ltqn1}
\end{equation}
and
\begin{equation}
\hat\psi_{\rm II}(u,s) = \int_0^\infty \left( \sum_{n=0}^\infty Q_n(z) s^n\right) e^{-u z} dz \ ,
\label{ltQn1l}
\end{equation}
have the explicit solutions~\cite{MCZ} 
\begin{equation}
\hat\psi_{\rm I}(u,s) = \frac{1}{u} \phi(0,s) \phi(u,s)
\label{ltqn2}
\end{equation}
and 
\begin{equation}
\hat\psi_{\rm II}(u,s) = \frac{1}{u}\left[ \frac{1}{u} \phi(0,s) -  \phi'(0,s) \right] \phi(u,s) \ ,
\label{ltQn2l}
\end{equation}
where the prime indicates differentiation with respect to $u$,
$\phi(0,s) = (1-s)^{-1/2}$, and
\begin{eqnarray}
\phi(u,s) &=& \frac{1}{\sqrt{1-s}+\sigma u \sqrt{s/2}} \nonumber \\
&\times&  \exp\left[-\frac{u}{\pi}
\int_0^{\infty}
\frac{dk}{u^2+k^2}\, \ln\left(\frac{1-s {\hat f}(k)}{1-s + s\sigma^2k^2/2}\right)\right] \ ,
\label{phius1}
\end{eqnarray}
where as before
${\hat f}(k)=\int_{-\infty}^{\infty} f(x) e^{ikx} dx$ is the Fourier transform of $f(x)$.

Using these solutions in Eq.\ (\ref{denprofn}) one can in principle calculate the full density profile
at all times $n$. In Ref.\ \cite{MCZ}, we have shown how to obtain the stationary solution (the $n\to \infty$
limit) explicitly. In the next section, we show how to obtain the two leading terms in the expression
for flux using the explicit solutions mentioned above.   

\section{Calculation of the discrete Flux}

In this section our aim is to compute the expression for the discrete flux, in particular
to prove the two leading terms in Eq.\ (\ref{rzflux}). To proceed, it turns out to be convenient
to compute not the instantaneous flux $\Phi(t)$ as in Eq.\ (\ref{rzflux}) but rather 
the 
cumulative flux $\Phi^{\rm{cum}}(t)$ up to time $t$.  We also replace $t$ by the subscript $n$ to 
reflect the discrete nature of time in this problem.  From (\ref{rzflux}) we expect
\begin{equation}
\Phi_n^{\rm{cum}} \sim 4 \pi (R - \epsilon) D \rho_0 
\left[ n + \frac{ 2(R - \epsilon)\sqrt{n}}{\sqrt{\pi D}} + {\cal O} (1)\right] \ ,
\label{totalflux}
\end{equation}
where the correction term ${\cal O} (1)$ is effectively the constant of integration.

To prove the result in Eq.\ (\ref{totalflux}), we note that 
the cumulative flux up to time $n$ is simply the total number of particles
absorbed by the sphere up to time $n$ and is given precisely by the 
``missing" mass at that time:
\begin{eqnarray}
\Phi_n^{\rm{cum}} &=& \int_R^\infty  4 \pi r^2 [\rho_0 - \rho_n(r)] dr \nonumber \\
 &=& 4 \pi  \rho_0 \int_R^\infty [r^2 - rF_n(r)] dr \nonumber \\
 &=& 4 \pi \rho_0 \int_0^\infty [(z + R)^2 - (z+R)(Q_n(z) + R q_n(z)) ] dz \ ,
\label{phin}
\end{eqnarray}
where in line two we used $F_n(r) = \rho_n(r)/(\rho_0 r)$ and in 
the third line we made the shift $z = r - R$
and used $F_n(z+R) = H_n(z) =(Q_n(z) + R q_n(z))$ as explained in the previous section.  Here $Q_n(z)$ 
and $q_n(z)$ are the solutions to Eqs.\ (\ref{Qn1l}) and (\ref{qn1}) whose Laplace transforms are 
given explicitly in Eqs.\ (\ref{ltQn2l}) and (\ref{ltqn2}) respectively. So, our job 
is to use these explicit solutions in the expression in Eq.\ (\ref{phin}) and obtain
the two leading terms for large $n$. The rest of the section is devoted precisely to this 
technical task.

Simplifying (\ref{phin}) further, we find
\begin{eqnarray}
\Phi_n^{\rm{cum}} &=& \ 4 \pi \rho_0 \int_0^\infty (z + R)[z + R - (Q_n(z) + R q_n(z)) ] dz \nonumber \\
&=& \ 4 \pi \rho_0 \int_0^\infty (z + R)[(z - Q_n(z)) + R(1- q_n(z)) ] dz \nonumber \\
&=& \ 4 \pi \rho_0 [J^{(1)}_n + R (J^{(2)}_n + J^{(3)}_n) + R^2 J^{(4)}_n]  \ ,
\label{phin1}
\end{eqnarray}
where
\begin{equation}
 J^{(1)}_n = \int_0^\infty z(z - Q_n(z))  dz \ ,
\end{equation}

\begin{equation}
 J^{(2)}_n =  \int_0^\infty (z - Q_n(z))dz  \ ,
\end{equation}

\begin{equation}
 J^{(3)}_n =  \int_0^\infty z(1 - q_n(z)) dz  \ ,
\end{equation}

\begin{equation}
 J^{(4)}_n =   \int_0^\infty (1 - q_n(z))  dz \ .
\label{J4def}
\end{equation}

One can verify that, for the case of the 
Rayleigh flight of jump length $\ell$ (\ref{uniform}), the 
instantaneous flux $\Phi_n = \Phi_n^{\rm cum} - \Phi_{n-1}^{\rm cum}$
calculated from the above formulas agrees with Eqs.\ (13) of Ref.\ 
\cite{Ziff1}:
\begin{equation}
\Phi_n = 4 \pi R D \rho_0 [a^{(1)}(n) - (\ell/R) b^{(1)}(n) 
+(R/\ell) a^{(2)}(n) - b^{(2)}(n) ]
\label{jspflux}
\end{equation} 
where
\begin{eqnarray} a^{(1)}(n) &=& 3 \int_0^1 (1-z) Q_n(z) dz  \ ,\qquad
 b^{(1)}(n) = \frac{3}{2} \int_0^1 (1-z)^2 Q_n(z) dz \ , \nonumber \\
 a^{(2)}(n) &=& 3 \int_0^1 (1-z) q_n(z) dz \ , \qquad
 b^{(2)}(n) = \frac{3}{2} \int_0^1 (1-z)^2 q_n(z) dz  \ .
 \label{asandbs}
 \end{eqnarray}

Now, to calculate the integrals $J_n^{(k)}$ for a general
jump-length distribution, we proceed as follows. We first Taylor-expand 
$\phi(u,s)$ defined in Eq.\ (\ref{phius1}) in powers of $u$ for small $u$. This gives
\begin{eqnarray}
\phi(u,s) &=& 
\frac{\phi(0,s)}{1 + a(s) u} e^{- u I_2(s) + u^3 I_4(s) + \ldots }=\phi(0,s) \bigg[ 1- \nonumber \\
& & [a(s) + I_2(s)]u + [a(s)^2 + a(s) I_2(s) + \frac 1 2 I_2(s)^2]u^2 + \ldots \bigg] \ ,
\label{expand}
\end{eqnarray}
where 
\begin{eqnarray}
\phi(0,s)&=& (1-s)^{-1/2} \ ,\\ 
a(s) &=& \frac{\sigma \sqrt{s}}{\sqrt{2(1-s)}} \ , \\
I_n(s) &=& \frac{1}{\pi}\int_0^{\infty} \frac{dk}{k^n}\, \ln\left(\frac{1-s {\hat f}(k)}{1-s + 
s\sigma^2k^2/2}\right)\ ,\quad\quad n\le 4 \ .
\label{In}
\end{eqnarray}
Note that the analytic expansion in small $u$ in the first line of Eq. (\ref{expand}) is valid up to
$O(u^3)$ and the $...$ refers to non-analytic higher order terms. For our purpose, we need terms only
up to $O(u^3)$.

Next, we carry out a Taylor-series expansion of $\hat\psi_{\rm I}$ and $\hat\psi_{\rm II}$ in powers of $u$:
\begin{eqnarray}
\hat\psi_{\rm I}(u,s) &=& 
\frac{1}{u(1-s)} - \frac{a(s)+I_2(s)}{1-s} \nonumber \\
&+& \frac{[a(s)^2 + a(s) I_2(s) + \frac{1}{2} I_2(s)^2 ]u}{1-s} + \ldots
\label{psiITaylor}
\end{eqnarray}

\begin{eqnarray}
\hat\psi_{\rm II}(u,s) &=& \frac{1}{u^2(1-s)} - \frac{a(s) I_2(s)+ \frac{1}{2} I_2(s)^2}{1-s} 
\nonumber \\
&+& \frac{[a(s)^2 I_2(s)+ a(s) I_2(s)^2 + \frac{1}{3} I_2(s)^3 + I_4(s)]u}{1-s} + \ldots
\label{psiIITaylor}
\end{eqnarray}
and rewrite $\hat\psi_{\rm I}$ and $\hat\psi_{\rm II}$ in a way that separates off the singular
part in $u$:
\begin{equation}
\hat\psi_{\rm I}(u,s) = \frac{1}{u(1-s)} - \sum_{n=0}^\infty s^n \int_0^\infty(1 - q_n(z)) e^{-u z} dz
\end{equation} 
\begin{equation}
\hat\psi_{\rm II}(u,s) = \frac{1}{u^2(1-s)} - \sum_{n=0}^\infty s^n \int_0^\infty(z - Q_n(z)) e^{-u z} dz \ ,
\end{equation}
so that the integrals remain convergent when $u \to 0$.  Expanding the exponentials 
$e^{-u z}$ in powers
of $u$, we get directly the generating functions of the desired integrals $J_n^{(k)}$.
For example, $ \sum s^n J^{(4)}_n$ follows from  (minus) the coefficient of $u^0$ in $\hat\psi_{\rm I}(u,s)$.
Thus, from Eq.\ (\ref{psiITaylor}) we deduce
\begin{equation}
\sum_{n=0}^\infty s^n J^{(4)}_n = \frac{a(s)+ I_2(s)}{1-s} =  \frac{\sigma \sqrt{s}}{\sqrt{2} (1-s)^{3/2}} + \frac{I_2(s)}{1-s} \ .
\label{J4genfct}
\end{equation}
We extract the large-$n$ behaviour of $J_n^{(k)}$ by looking at the behaviour of
the generating function for $s \to 1$.   To carry out expansions
to higher order, we use (see Appendix)
\begin{equation}
\frac{\sqrt{s}}{(1-s)^{3/2}} = \sum_{n=0}^\infty C_n s^n, \qquad C_n \sim 2 \sqrt{\frac{n}{\pi}}  + \frac1{4 \sqrt{\pi n}}
\qquad \hbox{as } n \to \infty \ .
\label{series}
\end{equation}
We also show in the Appendix that
\begin{equation}
I_2(s)  \sim I_2(1) -  \frac{ \sqrt{2} \mu_4}{24 \sigma^3} (1-s)^{1/2} + {\cal O}(1-s) \ ,
\label{I2expan}
\end{equation}
and
\begin{equation}
I_4(s) \sim  -\frac{  \mu_4}{24 \sqrt{2}\sigma} (1-s)^{-1/2} + {\cal O}(1) \ ,
\label{I4expan}
\end{equation}
as $s \to 1$, where
 $\mu_4 = \int_{-\infty}^{\infty} x^4 f(x) dx$  is the fourth moment of 
the projected jump distribution $f(x)$.

Putting these results into (\ref{J4genfct}), and using the
Taylor-series expansions of $(1-s)^{1/2}$ and $(1-s)^{-1/2}$,
we find
\begin{equation} 
J^{(4)}_n \sim \sigma \sqrt{\frac{2 n}{\pi}} - \epsilon+ \frac{\mu_4 + 3 \sigma^4}{12 \sigma^3} \sqrt{\frac1{2 \pi n}} + 
{\cal O}(n^{-1}) \ ,
\label{J4final}
\end{equation}
where we have used $\epsilon = -I_2(1)$.  

Likewise,  $\sum s^n J^{(3)}_n$ follows from the coefficient of 
$u^1$ in $\Psi_{\rm I}(u,s)$:
\begin{eqnarray}
\sum_{n=0}^\infty s^n J^{(3)}_n
&=& \frac{a(s)^2 + a(s) I_2(s) + \frac{1}{2} I_2(s)^2}{1-s} \nonumber \\
&=&  \frac{\sigma^2 s}{2 (1-s)^2}
+ \frac{\sigma \sqrt{s} I_2(s)}{\sqrt{2} (1-s)^{3/2}} + \frac{ \frac{1}{2} I_2(s)^2}{1-s} \ ,
 \label{J3genfct}
\end{eqnarray}
and using  (\ref{I2expan}), we find,
\begin{equation}
J^{(3)}_n \sim \frac{\sigma^2 n}2 - \sigma \epsilon \sqrt{\frac{2 n}{\pi}}
+ \frac{1}{2}\epsilon^2
- \frac{\mu_4}{24 \sigma^2} + {\cal O}(n^{-1/2}) \ .
\label{J3final}
\end{equation}
In the same way,  $\sum s^n J^{(2)}_n$ follows from the coefficient of 
$u^0$ in $\Psi_{\rm II}(u,s)$:
\begin{equation}
\sum_{n=0}^\infty s^n J^{(2)}_n = \frac{a(s) I_2(s) + \frac 1 2 I_2(s)^2}{1-s}
 = \frac{\sigma \sqrt{s} I_2(s)}{\sqrt{2} (1-s)^{3/2}} +\frac{  \frac{1}{2} I_2(s)^2}{1-s} \ ,
 \label{J2genfct}
 \end{equation}
which implies 
\begin{equation}
J^{(2)}_n \sim - \sigma \epsilon \sqrt{\frac{2 n}{\pi}}  +\frac{1}{2}\epsilon^2
- \frac{\mu_4}{24 \sigma^2} + {\cal O}(n^{-1/2}) \ .
\label{J2final}
\end{equation}
Finally,  $\sum s^n J^{(1)}_n$ follows from the coefficient of 
$u^1$ in $\Psi_{\rm II}(u,s)$:
\begin{eqnarray}
\sum_{n=0}^\infty s^n J^{(1)}_n &=& \frac{a(s)^2 I_2(s) + a(s) I_2(s)^2 + \frac{1}{3} I_2(s)^3 + I_4(s)}{1-s} \nonumber \\
 &=&   \frac{\sigma^2 s I_2(s)}{2 (1-s)^2} + \frac{\sigma \sqrt{s} I_2(s)^2}{\sqrt{2} (1-s)^{3/2}} + \frac{\frac{1}{3} I_2(s)^3}{1-s} + \frac{I_4(s)}{1-s} \ .
 \label{J1exp}
 \end{eqnarray}
Note that the first term on the RHS above behaves as 
\begin{equation}
  \frac{\sigma^2 s I_2(s)}{2 (1-s)^2} \sim \frac{\sigma^2 s I_2(1)}{2 (1-s)^2} + \frac{s \sqrt{2} \mu_4}{24 \sigma (1-s)^{3/2}} \ ,
  \end{equation}
and the last term of Eq.\ (\ref{J1exp}) behaves as
\begin{equation}
\frac{I_4(s)}{1-s} \sim - \frac{s \sqrt{2} \mu_4}{24 \sigma (1-s)^{3/2}} \ ,
\end{equation}
as $s \to 1$,
so between these two expressions, the terms containing $\mu_4$ 
exactly cancel out.
With the remaining terms of (\ref{J1exp}), we find
\begin{equation}
J^{(1)}_n \sim \frac{\sigma^2 \epsilon n}2
+ \sigma \epsilon^2 \sqrt{\frac{2 n}{\pi}} + {\cal O}(1) \ .
\label{J1final}
\end{equation}
The $I_2(s)^3$ term in Eq.\ (\ref{J1exp})
does not contribute to this order.

Finally, putting the four results Eqs.\ (\ref{J4final}), (\ref{J3final}), (\ref{J2final}) and 
(\ref{J1final}) into Eq.\ (\ref{phin1}), we find, after some algebra,
\begin{eqnarray}
\Phi_n^{\rm{cum}} &=& \ 4 \pi \rho_0 [ J^{(1)}_n + R(J^{(2)}_n+J^{(3)}_n) + R^2 J^{(4)}_n] \nonumber \\
&=&  \ 2 \pi \rho_0  (R-\epsilon) \sigma^2 \left[n + 2(R-\epsilon) \sqrt{\frac{2n}{\pi \sigma^2}}
 + {\cal O}(1) \right]\ .
 \label{final}
\end{eqnarray}
which is precisely in the form of Eq.\ (\ref{totalflux}).
Then $\Phi_n$ follows from $\Phi_n = \Phi_n^{\rm{cum}} -\Phi_{n-1}^{\rm{cum}}$, 
and the result is exactly in the form of 
Eq.\ (\ref{rzflux})
with $D = \sigma^2/2$.  This is the expected form of $D$ given that
$\sigma$ is the standard deviation of the one-dimensional projection 
of the three-dimensional walk.  (For example, for the Rayleigh flight, this
corresponds to the familiar formula $D = \ell^2/6$ \cite{Chandra}.) \ 
This completes the proof of our main result, Eq.\ (\ref{rzflux}).

We close this section with the following observation:  If we take the difference between 
(\ref{J3genfct}) and (\ref{J2genfct}), we find

\begin{equation}
\sum_{n=0}^\infty s^n (J^{(3)}_n-J^{(2)}_n) 
=\frac{\sigma^2 s}{2(1-s)^2}  \ ,
 \end{equation}
 which implies 
\begin{equation}
J^{(3)}_n-J^{(2)}_n = 
 \int_0^\infty (Q_n(z)  - z q_n(z))dz  = \frac{n \sigma^2}{2} \ ,
 \label{J32identity}
 \end{equation}
 exactly for all $n$.  For the Rayleigh jump case, the rhs is $n \ell^2/6$, and in that case
 this is the cumulative equivalent to the identity $a^{(1)}(n) = 1 - b^{(2)}(n)$ observed
 numerically in Ref.\ \cite{Ziff1}.
 Thus, we have proven that result, and shown that it can be generalized for an arbitrary
 jump distribution, with the difference given in Eq.\ (\ref{J32identity})
  depending only upon the standard deviation  $\sigma$ of the jump distribution.

\section{One dimension}
 
Let us now consider a one-dimensional system where the origin represents a trap and initially
we have particles distributed uniformly over the positive line $z>0$ with density $\rho_0$.
Each particle jumps at every time step and the jump distance $u$ is drawn from a symmetric
jump distribution $f(u)$. Whenever a particle crosses over the origin to the negative side
it gets trapped by the origin. We ask similar questions as in $d=3$: What is the density profile 
$\rho_n(z)$ at step $n$ and what is the instantaneous flux $\Phi_n$ at step $n$?

Due to the recurrent nature of the random walks in $d=1$, it is clear that eventually all the particles
will be absorbed by the origin and hence $\rho_{\infty}(z)=0$ for all $z\ge 0$. This is a
major difference from the 3-d case. In 1-d, the density profile for $z>0$ evolves in time via the integral 
equation
\begin{equation}
\rho_n(z) = \int_0^{\infty} \rho_{n-1}(z') f(z-z') dz' 
\quad {\hbox{with}} \quad
\rho_0(z)=\rho_0 \ .
\label{1dden1}
\end{equation}
Clearly, $\rho_n(z)= \rho_0 q_n(z)$ where $q_n(z)$ satisfies the integral equation (\ref{qn1}) with
the initial condition $q_0(z)=1$ for all $z>0$, and we already know its solution from section II.
Note that $q_n(z)$ can also be interpreted~\cite{CM} as the
probability that a random walker starting at $z>0$ and undergoing discrete jumps drawn from $f(z)$ does not cross 
the origin during the first $n$ steps, as discussed in Ref.\ \cite{CM}.

Regarding the flux, the cumulative amount $\Phi_n^{\rm{cum}}$ is given by the missing mass
at step $n$
\begin{equation}
\Phi_n^{\rm{cum}}= \int_0^{\infty}[\rho_0-\rho_n(z)]dz= \rho_0\int_0^{\infty}[1-q_n(z)]dz \ .
\label{1dcumflux1}
\end{equation}
Now, the integral $\int_0^{\infty} [1-q_n(z)]dz $ is precisely $J^{(4)}_n$ of Eq.\ (\ref{J4def})
whose asymptotic behaviour is given in Eq.\ (\ref{J4final}). The integral
$\int_0^{\infty} [1-q_n(z)]dz $ also represents
the expected maximum $E[M_n]$ up to $n$ steps of a random walker
starting at the origin and undergoing jumps drawn from $f(z)$ at every time step~\cite{CM}.
Thus, we have a rather nice relationship between the cumulative flux and the expected maximum,
\begin{equation}
\Phi_n^{\rm{cum}} = \rho_0 E[M_n] = \rho_0\left[\sigma \sqrt{\frac{2 n}{\pi}} - \epsilon + \frac{\mu_4 + 3 
\sigma^4}{12 \sigma^3} \sqrt{\frac1{2 \pi n}} +   {\cal O}(n^{-1})\right].
\label{1dcumflux2}
\end{equation}
where $\epsilon $ is the same quantity in Eq.\ (\ref{stat5}) that  appears in the expression for cumulative flux in three dimensions. However, while in  $3$-d,  $\epsilon$
retains its appearence  in the expression for the
instantaneous flux $\Phi_n = \Phi_n^{\rm{cum}} - \Phi_{n-1}^{\rm{cum}}$,
in $1$-d it disappears from  $\Phi_n$,
at least up to this order.

That is, from Eq.\ (\ref{1dcumflux2}), we have
\begin{eqnarray}
\Phi_n &=& \Phi_n^{\rm{cum}} - \Phi_{n-1}^{\rm{cum}} \nonumber \\
&=& \rho_0 \left(\sigma \sqrt{\frac{2}{\pi}}  \left[n^{1/2} - (n-1)^{1/2} \right] 
+  \frac{\mu_4 + 3  \sigma^4}{12 \sigma^3 \sqrt{{2 \pi}} } 
\left[n^{-1/2} - (n-1)^{-1/2} \right] \right)\nonumber \\
 &=& \rho_0 \sigma \sqrt{\frac{2}{\pi}} \bigg(   n^{1/2}  \left[ 1 -  \left(1 - \frac1{2n} - \frac1{8n^2}+ \ldots\right)\right]  \nonumber \\
 & &  \qquad +  \frac{\mu_4 + 3 \sigma^4 }{24 \sigma^4} n^{-1/2}  \left[1 - \left(1 +  \frac1{2n} + \ldots\right)\right] \bigg)  \nonumber \\
  &=& \rho_0 \frac{ \sigma}{\sqrt{{2}{\pi}}}\left(   n^{-1/2}  +  \frac{3 \sigma^4 - \mu_4 }{24 \sigma^4}\,  n^{-3/2}  
  +\ldots \right)
  \label{phingeneral}
   \end{eqnarray}
 which does not involve $\epsilon$.  Note that the coefficient of the last term
 is proportional to the ``excess," defined as
 $\gamma_2 = \mu_4/\sigma^4-3$.  This quantity equals zero for a gaussian distribution.
 
 For the case of the uniform $f(x)$ of Eq.\  (\ref{uniform}),
we have $\sigma^2 = \ell^2/3$ and
 $\mu_4 = \ell^4/5$, and from Eq.\ (\ref{phingeneral}), it follows that the flux
 is given by
\begin{equation}
\Phi_n= \rho_0 \ell \sqrt{\frac{1}{6 \pi}} \left( n^{-1/2}  +  \frac{1}{20 } n^{-3/2}  
  +\ldots \right) \ .
  \label{phinrayleigh}
 \end{equation}
This result proves one more 
formula that was conjectured in Ref.\ \cite{Ziff1}.
For the uniform $f(x)$,  $\Phi_n$ equals the 
 $\rho_0 \ell a^{(2)}(n)/6$ where $a^{(2)}(n)/6$ is given in  Eq.\ (\ref{asandbs}),
and in Eq.\ (21c) of Ref.\ \cite{Ziff1}
is was conjectured from the numerical results that 
\begin{equation}
a^{(2)}(n)/6 = (6 \pi)^{-1/2} n^{-1/2} + (1/20) (6 \pi)^{-1/2} n^{-3/2} + {\cal O}(n^{-5/2}) \ .
\label{a2}
\end{equation}
Indeed, the above formula is identical to Eq.\ (\ref{phinrayleigh}).
Thus, we have proven Eq.\ (\ref{a2}),
and in particular verified that the coefficient in the second term
 is exactly 1/20.

Note, finally, for the uniform $f(x)$, 
Eq.\ (\ref{1dcumflux2}) becomes  
\begin{equation}
\Phi_n^{\rm{cum}} = \rho_0 \ell
\left[\sqrt{\frac{2}{3 \pi}}n^{1/2} - c + \frac1{5} \sqrt{\frac{2}{3 \pi}}n^{-1/2} + 
{\cal O}(n^{-1})\right]
\label{onedflux}
\end{equation}
where $c=0.29795219\dots$ as given in Eq.\ (\ref{c2}).
This result agrees with the expectation of the maximum of a 
one-dimensional walker with uniform jump density~\cite{CFFH,CM}, $E[M_n]$, which
is identical to $\Phi_n^{\rm{cum}}$ (up to the constant factor $\rho_0$) as mentioned above.

\section{Conclusions}
Thus, we have proven that Eq.\ (\ref{rzflux}) or its integrated form
Eq.\ (\ref{totalflux}) applies to an arbitrary jump distribution
$f(x)$.  Here $f(x)$ is related to the radial projection of the 
three-dimensional jump distribution $W(|\vec r-\vec r'|)$ by
Eq.\ (\ref{jump}) as long
as $W(u)$ has a cutoff for large $u$.  The generality of the flux
in Eq.\ (\ref{rzflux}) implies that there is a universal long-time 
behaviour of the intermediate particle density given by
\begin{equation}
\rho(r,t) \sim \rho_0 \left( \frac{r - R + \epsilon}{r}\right) \left(1 + \frac{R-\epsilon}{\sqrt{\pi D t}}\right) \ ,
\label{density}
\end{equation}
valid for large $t$ and $R \ll r \ll \sqrt{\pi D t}$.  Only the length $\epsilon$ and diffusion
coefficient $D$ depend upon
the details of the jump distribution.  Eq.\ (\ref{density}) is equivalent to Eq.\ (\ref{denprof1})
with $R$ replaced by $R - \epsilon$ and the complementary error function expanded to two terms.

The integrated form of the flux of 
Eq.\  (\ref{totalflux}) brings in additional
constants that do not appear in the instantaneous form, Eq.\ (\ref{rzflux}),
including the quantity $\epsilon$ that appears in the one-dimensional
flux, Eq.\ (\ref{onedflux}).  Indeed, 
$\epsilon$ defined in Eq.\ (\ref{stat5}) is ubiquitous as it
appears in
many terms in the expansions of the integrals $J_n^{(k)}$ that apply
to both one- and three-dimensional problems.

The integrated form of the flux also
enters directly  in the formula for the trap survival probability  (the probability
that an adsorbing sphere is not hit by a diffusing particle up to time $t$):
$S(t) = \exp[-\Phi^{\rm cum}(t)]$.  This quantity has been studied
extensively and has applications to quenching of fluorescence 
reactions (e.g., \cite{ZhouSzabo,DongAndre,IbukiUeno}).

When the jump probability corresponds to the Rayleigh
flight, $f(x)$ becomes the uniform jump distribution of Eq.\ (\ref{uniform}), in which case
$\epsilon$ becomes  $0.29795219\ldots \ell$ found previously \cite{Ziff1,MCZ}.
For this model
 we have also provided proofs of the conjectured behaviour of the quantities
$a^{(1)}(n)$, $b^{(1)}(n)$, $a^{(2)}(n)$, and $b^{(2)}(n)$ in Eqs.\ (21a)--(21d) of Ref.\ \cite{Ziff1}
and proven the conjectured identity $a^{(1)}(n) = 1 - b^{(2)}(n)$.  We did not 
derive expressions for the higher-order corrections of Eq.\ (21b) and Eq.\ (21d) 
of Ref.\ \cite{Ziff1} that involve
the empirically determined
constant $d  = 0.0270103$, but these terms enter only in higher-order 
corrections to the flux.

We also proved the identity (\ref{J32identity}) for a general jump distribution.
This identity relates the solution $q_n(z)$ with a linear initial condition
to the solution $Q_n(z)$ with a constant initial condition, and implies a
subtle relation between the
corresponding one- and three-dimensional diffusion
processes.  

In this paper we have restricted ourselves to the spatial dimensions $1$ and $3$. A natural
question is: what are the expressions for the density profile $\rho_n(r)$ and the cumulative flux
$\Phi_n^{\rm{cum}}$ in dimensions $d\ne 1, 3$? Is it possible to compute 
the Milne extrapolation length $\epsilon$ for $d>3$?  
The reason we restricted ourselves only to $d=1$ and $d=3$
can be traced back to the basic integral equation (\ref{evol2}). For $d=1$ and
$d=3$, this integral equation can be reduced to a one-dimensional integral equation
(\ref{evol6}) with a stationary kernel $f(z-z')$ that is symmetric. For $d\ne 1, 3$
this step cannot be done and finding an exact solution of the integral equation
seems difficult. For example, in $d=2$, using $d\vec r'= r' dr' d\theta$ in Eq.\ (\ref{evol2})
one gets
\begin{equation}
\rho_n(r) = \int_R^{\infty} dr'\, \rho_{n-1}(r')\,r' \int_0^{2\pi} 
W\left(\sqrt{r^2+r'^2-2rr'\cos{\theta}}\right)\, d\theta.
\label{evol2d}
\end{equation}
If one performs the angular integration after the substitution $z^2=r^2+r'^2-2rr'\cos \theta$
one gets the following integral equation
\begin{equation}
\rho_n(r) =\int_R^{\infty} dr'\,   \rho_{n-1}(r')\, G(r,r')
\label{evol2d1}
\end{equation}
where the kernel $G(r,r')$ is a rather complicated function of both $r$ and $r'$  
\begin{equation}
G(r,r') = 2r' \int_{|r-r'|}^{r+r'}\frac{dz\, z\, W(z)}{\sqrt{4r^2r'^2-(z^2-r^2-r'^2)^2}} \ .
\label{evol2d2}
\end{equation}
Solving the integral equation (\ref{evol2d1}) or the equivalent integral equations for $d>3$ remain a 
challenging problem.   

\vspace{0.5cm}

\appendix
\section{Proof of Eqs.\ (\ref{series}) and (\ref{I2expan}) }
Proof of Eq.\ (\ref{series}):
For the terms that involve $a(s)/(1-s)$, we have for $s$ near the singular
point at $s = 1$,
\begin{eqnarray}
\frac{\sqrt{s}}{(1-s)^{3/2}} &=& \frac{(1 - (1-s))^{1/2}}{(1-s)^{3/2}} \nonumber \\
&\sim&  (1-s)^{-3/2} - \frac1 2 (1-s)^{-1/2} \nonumber \\
&=& \sum_{n=0}^\infty (2n + 1) {2n \choose n} \frac{s^n}{2^{2n}}  - \frac1 2\sum_{n=0}^\infty {2n \choose n} \frac{s^n}{2^{2n}}  \nonumber \\
&=& \sum_{n=0}^\infty C_n s^n \ .
\end{eqnarray}
Now, for large $n$, 
\begin{equation}
\frac{1}{2^{2n}}{2n \choose n} \sim \sqrt{\frac{1}{\pi n}} \left[ 1 - \frac{1}{8n} + \ldots \right] \ .
\end{equation}
Therefore, we have
\begin{equation}
C_n \sim \frac1{\sqrt{\pi n}} \left[ 1 - \frac1{8n} + \ldots \right] (2 n + 1 - \frac 1 2)
= 2\sqrt{\frac n \pi}  + \frac{1}{4 \sqrt{\pi n}} + \ldots \ ,
\end{equation}
thus contributing to the $n^{1/2}$ and $n^{-{1/2}}$ terms in the large-$n$
behaviour.

Derivation of Eq.\ (\ref{I2expan}):
First differentiate Eq.\ (\ref{In}) (for $n = 2$)
 with respect to
$s$, then put $s=1-\varepsilon$ where $\varepsilon$ is small. Next, inside the
corresponding integrand make the change of variable $k=\sqrt{\varepsilon} y$
and then expand everything for small $\varepsilon$.
Use ${\hat f}(k) \sim 1-\sigma^2 k^2/2 + \mu_4 k^4/24 .$
What you get as $\varepsilon\to 
0$ is the following:
\begin{equation}
\frac{dI_2}{d{\varepsilon}} \sim \frac{\mu_4 \sqrt{2} }{12 
\pi \sigma^3} \varepsilon^{-1/2} \int_0^{\infty} \frac{z^2 dz}{(1+z^2)^2} \ .
\end{equation}
The integral gives $\pi/4$. Then integrating the above with 
respect to $\varepsilon$ one gets
\begin{equation}
I_4(\varepsilon) \sim I_4(\varepsilon=0) + C_1 \sqrt{\varepsilon} \ ,
\end{equation}
where $C_1=\sqrt{2} \mu_4/(24 \sigma^3)$.

For $I_4$, the integral is already singular as $\varepsilon \to 0$ and
there is no need to differentiate.  A direct analysis of the integral
gives Eq.\ (\ref{I4expan}).

\ack

R.M.Z. acknowledges support from NSF grant DMS-0553487.

\end{document}